\documentstyle[12pt]{article}
\newcommand{\be}[1]{\begin{equation} \label{(#1)}}
\newcommand{\ee}{\end{equation}}
\newcommand{\ba}[1]{\begin{eqnarray} \label{(#1)}}
\newcommand{\ea}{\end{eqnarray}}

\newcommand{\rf}[1]{(\ref{(#1)})}
\def \lsim {\mbox{${}^< \hspace*{-7pt} _\sim$}}
\def \gsim {\mbox{${}^> \hspace*{-7pt} _\sim$}}
\def\p{\prime}
\def\rp{$R_p \hspace{-1em}/\ \  $}
\def\rpm{R_p \hspace{-0.8em}/\;\:}
\def \sw {\sin\!\theta^{}_W }
\def \cw {\cos\!\theta^{}_W }

\def \sbt {\sin\!\beta }
\def \cbt {\cos\!\beta }
\def \lg  {\langle}
\def \rg  {\rangle}
\def \tbt {\tan\!\beta }
\def \znbb {0\nu\beta\beta}
\begin{document}

\begin{center}
   {\bf Super-Kamiokande Constraints on R-parity
	Violating Supersymmetry}\\[3mm]
 	Vadim Bednyakov${}^{2}$, Amand Faessler${}^{1}$,
	and Sergey Kovalenko${}^{1,2}$ \\[1mm]
${}^{1}${\it Institute f\"ur Theoretische Physik  der Universit\"at
T\"ubingen,
Auf der Morgenstelle 14, D-72076 T\"ubingen, Germany}\\[0pt]
${}^{2}${\it Joint Institute for Nuclear Research, Dubna, Russia}
\end{center}
\bigskip

\begin{abstract}
We consider the neutrino oscillations within the minimal
supersymmetric standard model with R-parity violation (\rp MSSM).
The Super-Kamiokande atmospheric neutrino data are used to
set limits on the bilinear R-parity violating terms of the \rp MSSM.
These very stringent limits are out of reach of the other
experiments at present and in the near future.
\end{abstract}
\vskip 0.5cm
	The neutrino masses and mixing constitute one
	of the most pressing problem of particle physics.
	At present, there is a convincing experimental evidence for
	a non-trivial structure of the three-generation neutrino
	mass matrix.
	It derives from the solar
\cite{solar-1} and atmospheric
\cite{atm-1} neutrino deficit as well as, likely, from the LSND
	experiment
\cite{LSND}.
	All the three types of observations can be explained by
	the (anti)neutrino oscillations.
	An experimental breakthrough
	in exploration of the atmospheric neutrino anomaly has
	been recently made by the Super-Kamiokande collaboration
\cite{SuperK}
	measured the zenith angle dependence of the atmospheric
	neutrino flux.
	This high statistics experiment confirmed that the most likely
	solution of the atmospheric neutrino anomaly implies
	neutrino oscillation preferably in
	$\nu_{\mu} \rightarrow \nu_{\tau}$ channel
\cite{SuperK, bww98} or,
	maybe, partially in active-sterile channel
	$\nu_{\mu} \rightarrow \nu_{s}$
\cite{gnpv98}.
	This result may have
	far reaching consequences for the new physics phenomenology
\cite{nu-mass-theory}.

	From the theoretical side the neutrino mass problem
	has been attracting many efforts during all the history of
	modern particle physics. It stimulated construction of many
	interesting approaches
	which have their meaning far beyond the study of the neutrino
	properties.
	The presently popular supersymmetric (SUSY) models with R-parity
	non-conservation (\rp SUSY) may provide us with a new insight
	into the neutrino mass problem.
	In this framework neutrinos can acquire the tree-level
	supersymmetric masses via the mixing with the gauginos and
	higgsinos at the weak-scale
\cite{nu-LH, bgnn96}
	which is possible due to the presence of the bilinear R-parity
	violating operators in the superpotential and soft
	SUSY breaking sector.
	Remarkable, that this mechanism does not involve the physics at
	the large energy scales $M_{int} \sim {\cal O}(10^{12}$~GeV),
	in contrast to the see-saw mechanism,
	but relates the neutrino properties to the weak-scale
	physics accessible for the experimental studies.

	In the present paper we consider an impact of the new atmospheric
	neutrino data on the bilinear R-parity violation.
	We use the constraints
	on neutrino oscillation parameters recently obtained in Ref.
\cite{bww98}
	from the combined fit to the Super-Kamiokande atmospheric
\cite{SuperK} and to the CHOOZ reactor neutrino data
\cite{CHOOZ}.
	On this basis we derive new stringent constraints on
	the R-parity violating parameters of all the three generations.
	These constraints may have important implications for
	the lepton number/flavor violating processes such as
	the neutrinoless double beta ($\znbb$) decay,
	muon--electron(positron) conversion, etc.
	The \rp SUSY contribution to these processes can be expressed in
	terms of the quantities observed in the neutrino oscillation
	experiments. We have found that
	the Super-Kamiokande casts significantly more stringent limits
	on the 1st generation bilinear R-parity
	violation than the present $\znbb$-decay experiments.
	Previously, before the Super-Kamiokande data have been published,
	$\znbb$-decay experiments provided the
	best limits of this type
\cite{FKS97}.

	Let us consider the neutrino oscillations in
	the minimal supersymmetric standard
	model with R-parity violation (\rp MSSM).
	Recall that the R-parity is a multiplicative $Z_2$
	symmetry defined as
	$R_p=(-1)^{3B+L+2S}$, where $S,\ B$ and $L$ are the spin,
	the baryon and the lepton quantum numbers.

	The $R_p$ violation is introduced into the theory through
	the superpotential
\ba{W_rp}
W_{\rpm} = \lambda_{ijk}L_i L_j E^c_k +
	   \lambda^{\p}_{ijk}L_i Q_j D^c_k + \mu_j L_j H_2 +
	   \lambda^{\prime\prime}_{ijk} U^c_i D^c_j D^c_k,
\ea
	and soft SUSY breaking scalar potential terms
\ba{V_rp}
V_{\rpm}^{\rm soft} =
    \Lambda_{ijk}\tilde L_i \tilde L_j \tilde E_k^c
  + \Lambda^{\prime}_{ijk}\tilde L_i \tilde Q_j \tilde D_k^c
  + \Lambda^{\prime\prime}_{ijk}\tilde U_i^c \tilde D_j^c \tilde D_k^c
  + \tilde\mu_{2j}^2\tilde L_j H_2
  + \tilde\mu_{1j}^2\tilde L_j H^{\dagger}_1 + \mbox{H.c.}
\ea
	Here $L$,  $Q$  stand for lepton and quark
	doublet left-handed superfields while $ E^c, \  U^c,\   D^c$
	for lepton and {\em up}, {\em down} quark singlet  superfields;
	$H_1$ and $H_2$ are the Higgs doublet superfields
	with a weak hypercharge $Y=-1, \ +1$, respectively. Fields
	in Eq. \rf{V_rp} are the scalar components of these superfields.
	Summation over the generations is implied.
	To prevent fast proton decay it is a common practice to
	set $\lambda^{\prime\prime}= \Lambda^{\prime\prime}=0$.

	The impact of the R-parity violation on the low
	energy phenomenology is twofold.
	First, it leads to the lepton number/flavor violating
	interactions directly from the trilinear
	terms in $W_{\rpm}$.
	Second, bilinear terms in $W_{\rpm}$ and in
	$V_{\rpm}^{\rm soft}$ generate the non-zero vacuum
	expectation value for the sneutrino fields
	$\langle\tilde\nu_{i}\rangle\neq 0$ and cause
	neutrino--neutralino as well as electron--chargino mixing.
	The mixing gives rise to
	the non-trivial neutrino mass matrix and brings in the new lepton
	number/flavor violating interactions in
	the physical mass eigenstate basis.
	These interactions contribute to some exotic processes such as
	$\znbb$-decay, $\mu^- - e^{\pm}$ conversion, etc.
\cite{FKS97}
	and, therefore, can be constrained experimentally.
        In the present paper we analyze constraints
	on the neutrino mass matrix of the \rp MSSM from the
	neutrino oscillation experiments.

	Let us recall the mechanism of neutrino mass generation
	in the \rp MSSM.
	The bilinear terms in the Eqs.
\rf{W_rp} and \rf{V_rp}
	lead to the terms in the scalar potential linear in the sneutrino
	fields $\tilde{\nu}_{i}$.
	As a result, at the minimum of the potential
	these fields acquire non-zero vacuum expectation values
	$\langle\tilde{\nu}_{i}\rangle\neq 0$. Thus, the MSSM vertices
	$\tilde{Z}\nu \tilde{\nu}$ and  $\tilde{W}e\tilde{\nu}$
	create the gaugino--lepton mixing mass terms
	$\tilde{Z}\nu\langle\tilde{\nu}\rangle,
	\tilde{W}e\langle\tilde{\nu}\rangle$
	(with $\tilde{W},\tilde{Z}$ being wino and zino fields).
	Mixing of the leptons with the Higgsinos comes from
	the $\mu_{i} L_{i} H_2$ term in
	the superpotential
\rf{W_rp}.

	In the two component Weyl basis
\ba{basis1}
\Psi_{(0)}^{\p T}
&=& (\nu^{\p }_e,\, \nu^{\p }_{\mu},\, \nu^{\p }_{\tau},\,
     -i\lambda ,\, -i\lambda _{3} ,\,
      \tilde{H}^0_1,\, \tilde{H}_2^0),
\ea
	the mass term of the neutral fermions is
\ba{neut_L}
{\cal L}^{(0)}_{\rm mass}
	= -{1 \over 2}\Psi_{(0)}^{\p T}{\cal M}_0
                           \Psi^{\p}_{(0)}\ - \ \mbox{H.c.}\, ,
\ea
	Here $\nu'_e, \nu'_{\mu}, \nu'_{\tau}$ are the neutrino
	weak eigenstate fields, $\lambda$ and $\lambda_{3}$ are the
	$U_{1Y}$ and $SU_{2L}$ gauginos, respectively, while the
	higgsinos are denoted as $\tilde H^0_{1,2}$.

	The $7\times 7$ mass matrix has the see-saw structure
\ba{neut_m}
{\cal M}_0=\left(\begin{array}{cc}
	0   & m \\
	m^T & M_{\chi}
		\end{array}\right),
\ea
	with $3\times 4$ matrix
\ba{subm}
m=\left(\begin{array}{cccc}
-M_Z s_W c_{\beta} v_e      & M_Z c_W c_{\beta} v_e      &0& -\mu u_e \\
-M_Z s_W c_{\beta} v_{\mu}  & M_Z c_W c_{\beta} v_{\mu}  &0& -\mu u_{\mu} \\
-M_Z s_W c_{\beta} v_{\tau} & M_Z c_W c_{\beta} v_{\tau} &0& -\mu u_{\tau}
\end{array}\right),
\ea
	originating from the \rp bilinear terms in the superpotential
	\rf{W_rp} and the soft SUSY breaking sector \rf{V_rp}.
	Here
	$\tbt = \langle H_2^0\rangle/\langle H_1^0\rangle$,
	$s_W = \sw $, $c_W = \cw $,
	$s_{\beta} = \sbt$, $c_{\beta} = \cbt$.
	The supersymmetric Higgs mass $\mu$ occurs in
	$\mu H_1 H_2$ term of the MSSM superpotential
\cite{mssm}.
	We introduced the quantities
$v_{\alpha} = \langle\tilde\nu_{\alpha}\rangle/\langle H_1^0\rangle$,
$u_{\alpha} = \mu_{\alpha}/\mu$ which are assumed to be small,
$v_{\alpha}, u_{\alpha} << 1 $,
	and used as expansion parameters.
	For the details of this approximate approach we
	refer to Ref.
\cite{Now}.
	Hereafter $\alpha = e, \mu, \tau$ is a flavor index.
	In Eq. \rf{neut_m} $M_{\chi}$ is the usual $4 \times 4$ MSSM
	neutralino mass matrix in the basis
$\{-i\lambda, -i\lambda_3, \tilde{H_1^0}, \tilde{H_2^0} \}$
\ba{MSSM-chi}
M_{\chi} = \left( \begin{array}{cccc}
     M_1  &  0        & -M_Z s_W c_{\beta}  &  M_Z s_W s_{\beta}  \\
       0  &  M_2      &  M_Z c_W c_{\beta}  & -M_Z c_W s_{\beta}  \\
 -M_Z s_W c_{\beta}   &  M_Z c_W c_{\beta}  &  0     &  -\mu      \\
  M_Z s_W s_{\beta}   & -M_Z c_W s_{\beta}  & -\mu   &    0       \\
                   \end{array} \right).
\ea
	In the mass eigenstate basis defined as
\ba{mass_0}
\Psi_{(0)i} = \Xi_{ij} \Psi_{(0)j}^{\p},
\ea
	the $7\times 7$ neutral fermion mass matrix
	${\cal M}_0$ in Eq. \rf{neut_m} becomes diagonal
\ba{diag}
\Xi^*{\cal M}_0\Xi^{\dagger} = {\rm Diag} \{m_{\nu_i}, m_{\chi_k}\},
\ea
	where $m_{\nu_i}$ and $m_{\chi_i}$ are the physical neutrino
	and neutralino masses.
	For the considered case of the tree-level mass matrix
	only one neutrino has a non-zero mass
	$m_{\nu_1}=m_{\nu_2} =0,\
	m_{\nu_3} \neq 0$.
	This is a result of the minimal
	field content and the gauge invariance. The mass matrix
	${\cal M}_0$ in Eq. \rf{neut_m} has
	such a texture that its first three rows and the last one are
	linearly dependent and, therefore, two neutrino mass eigenstates
	are degenerate massless states.
	The $\nu_ {1}-\nu_{2}$ mass degeneracy is lifted by
	the 1-loop corrections as well as by
	the non-renormalizable terms in the superpotential.
	As a result $\nu_{1,2}$ acquire small non-equal masses
	$m_{\nu_1}\neq m_{\nu_{2}} \neq 0$
\cite{bgnn96}.
        This effect is not important for our analysis
	since we confine ourselves to the atmospheric
	neutrino anomaly explained as
	$\nu_{\mu}\rightarrow \nu_{\tau}$ oscillations.
        In this case the 1-loop corrections and non-renormalizable
        terms can be neglected as they give only small corrections to
        the tree level value of the corresponding oscillation parameter
        $\delta m^2_{\rm atm} =
        m^2_{\nu_{1,2}} - m^2_{\nu_{3}} =  m^2_{\nu_{3}} \neq 0$.
        This is not the case for the treatment of the solar neutrino data,
        interpreted in terms of $\nu_{e}\rightarrow \nu_{\mu}$ oscillations
        with $\delta m^2_{\rm sol} = m^2_{\nu_{1}} - m^2_{\nu_{2}} = 0$
        at tree level in the \rp MSSM.

	The mixing within the neutrino sector to leading order
	in the small expansion parameters $v_{\alpha}, u_{\alpha}$
	is described by
\ba{neutrino}
\nu_k = V^{(\nu)^*}_{\alpha k} \nu^{\p}_{\alpha}
\ea
	with $\nu^{\p}_{\alpha} =
	(\nu^{\p}_e, \nu^{\p}_{\mu}, \nu^{\p}_{\tau})$
	being the weak basis.

        The $3\times 3$ matrix $V^{(\nu)}$ diagonalizes
	the \rp-induced effective neutrino mass matrix:
\ba{nu-rotat}
V^{(\nu)^T}\; m_{\rm eff}\; V^{(\nu)}
        = {\rm Diag} \{0, \; 0, \; m_{\nu_{3}}\}.
\ea
	The tree-level expression for this mass matrix can
	be found in Ref.
\cite{Now}.
	The only non-zero eigenvalue which we identify with the
	tau neutrino mass  is given by
\ba{nu-tau-again}
m_{\nu_{3}}= F_{_{\rm MSSM}}\cdot
	\sum_{\alpha}\vert {\Lambda_{\alpha}} \vert^2\, .
\ea
	where
\ba{vec-again}
\Lambda_{\alpha} =
	\mu \lg\tilde\nu_{\alpha}\rg - \lg H_1\rg \mu_{\alpha}, \ \
F_{_{\rm MSSM}} =
	g_2^2 \left|\frac{M_1+\tan^2\theta_W M_2}
	{4\ {\rm Det}\ M_{\chi}}\right|.
\ea
	The function  $F_{_{\rm MSSM}}$ depends only on the
	R-parity conserving MSSM parameters.
	The determinant of the MSSM neutralino mass matrix
	\rf{MSSM-chi} is
\ba{detchi}
{\rm Det}\ M_{\chi} =
	\sin\!2\beta \ M_W^2 \ \mu (M_1 + \tan^2\theta_W  M_2)
	-  M_1 M_2 \mu^2.
\ea
     The \rp MSSM neutrino mixing matrix can be written in the following
form
%%%%%%%%%%%%%%%%%%%%%%%
\footnote{The \rp MSSM neutrino mixing matrix given in Ref. \cite{Now}
          contains misprints.}
%%%%%%%%%%%%%%%%%%%%%%%
\ba{nu-mix-exp}
V^{(\nu)}=\left(\begin{array}{ccc}
\cos \theta_{13} & 0 & -\sin \theta_{13} \\  %1
\sin \theta_{23}\sin \theta_{13} &
\cos \theta_{23} & \sin \theta_{23} \cos \theta_{13} \\ %2
\cos \theta_{23} \sin \theta_{13} & - \sin \theta_{23}
& \cos \theta_{13}\cos \theta_{23}        % 3
\end{array}\right) ,
\ea
	where the mixing angles are defined as follows:
\ba{tan-nu}
\tan \theta_{13} =
- \frac{\Lambda_e}{\sqrt{\Lambda_{\mu}^2 + \Lambda_{\tau}^2}}, \;\;\;\;\;
\tan \theta_{23} = \frac{\Lambda_{\mu}}{\Lambda_{\tau}} \, .
\ea
We used a freedom in definition of the mass eigenstates $\nu_{1}, \nu_{2}$
with $m_{\nu_1} = m_{\nu_2} = 0$ to set $V^{(\nu)}_{12} = 0$ in
Eq. \rf{nu-mix-exp}.
%
%%%%%%%%%%%%%%%%%%%%%%%%%%%%%%%%%%%%%%%%%%%%%%%%%%%%%%%

        Now, let us turn to neutrino oscillations in vacuum.
        The survival probability for a given
        neutrino flavor $\nu_{\alpha}$ is described
        by the formula
\cite{bwcp80}
\ba{surv}
P(\nu_{\alpha}\rightarrow \nu_{\alpha}) =
	1 - 4\sum_{i<j} P_{\alpha j} P_{\alpha i}
	\sin^2 \frac{\delta m^2_{ji}\cdot L}{4 E},
\ea
	where  $\delta m^2_{ji} = m^2_{\nu_j} - m^2_{\nu_i}$ and
	$P_{\alpha j} = \left|V^{(\nu)}_{\alpha j}\right|^2 $.

	From the combined fit to the Super-Kamiokande
\cite{SuperK} and the CHOOZ
\cite{CHOOZ} data the following constraints at 95\% C.L. were
	recently found in Ref.
\cite{bww98}:
\ba{fit}
	0.5 &\leq& \delta m^2_{\rm atm}/(10^{-3}{\rm~eV}^2) \leq 10.0 \,, \\
\label{(Pe3)}
	0.00 &\leq& P_{e3} \leq 0.08 \,, \\
	0.25 &\leq& P_{\mu 3} \leq 0.75 \,, \\
\label{(Ptau3)}
	0.25 &\leq& P_{\tau 3} \leq 0.75.
\ea
	The analysis of Ref.
\cite{bww98} was made under the assumption
	$|\delta m^2_{\rm atm}| >> |\delta m^2_{\rm sol}|$, with
	$\delta m^2_{\rm atm} = \delta m^2_{32}\approx \delta m^2_{31}$ and
	$\delta m^2_{\rm sol} = \delta m^2_{21}$.
	This covers a particular case of the \rp MSSM neutrino
	mass spectrum Eqs.
	\rf{nu-rotat} and \rf{nu-tau-again}
	which leads to the relations
\ba{mass2}
\delta m^2_{\rm atm} = \delta m^2_{32} =\delta m^2_{31} =
	m_{\nu_{3}}^2 = F_{_{\rm MSSM}}^2\cdot
	\left(\sum_{\alpha}\vert {\Lambda_{\alpha}} \vert^2
	\right)^2.
\ea
	and $\delta m^2_{\rm atm} >>
	\delta m^2_{\rm sol} = m_{\nu_{2}}^2 - m_{\nu_{1}}^2$
	since $m_{\nu_{1,2}}$ are zero at the tree level.

	The mixing parameters in Eqs.
	\rf{Pe3}--\rf{Ptau3} are determined by
	the tree level mixing matrix \rf{nu-mix-exp} as
\ba{Pij}
P_{\alpha 3} =
\frac{\Lambda_{\alpha}^2}{\sum_{\alpha}\vert {\Lambda_{\alpha}} \vert^2}.
\ea
	It is useful to introduce the effective SUSY parameters
\ba{Pdm}
{\cal A}_{\alpha} \equiv \Lambda_{\alpha}^2 \cdot F_{_{\rm MSSM}}  =
        P_{\alpha 3} \sqrt{\delta m^2_{\rm atm}} \ .
\ea
	An advantage of these parameters for our analysis follows from
	the fact that they depend on the \rp parameters
	$\lg\tilde\nu_{\alpha}\rg$, $\mu_{\alpha}$ from
	only one generation $\alpha$ unlike parameters $P_{\alpha 3}$,
	$\delta m^2_{atm}$, depending on
	the \rp parameters from different generations.

	The Eqs. \rf{fit}--\rf{Ptau3} cast the constraints on
	the effective parameters ${\cal A}_{\alpha}$:
\ba{eff-const}
0 \leq {\cal A}_{e}\leq 8\cdot 10^{-3}~{\rm eV}, \ \ \
5.6\cdot 10^{-3}~{\rm eV} \leq
{\cal A}_{\mu, \tau} \leq 7.5\cdot 10^{-2}~{\rm eV}.
\ea
	This constraint represents a complex exclusion condition
	imposed by the neutrino oscillation data on the \rp
	MSSM parameter space.

	It is instructive to obtain typical individual constraints on
	the superpotential parameters $\mu_{\alpha}$ and on the
	sneutrino vacuum expectation value $\lg\tilde\nu_{\alpha}\rg$.
	These constraints can be derived from Eqs.
\rf{eff-const} at typical weak scale values of
	the MSSM parameters $M_{1,2} \approx \mu \approx M_{_{\rm SUSY}}$
	and $\tbt = \lg H_2^0\rg/\lg H_1^0\rg = 1$.
	Here $M_{_{\rm SUSY}}$ is a characteristic SUSY breaking
	mass scale varying most likely in the interval
	$100~\mbox{GeV}~\lsim ~M_{_{\rm SUSY}}
	~\lsim ~1~\mbox{TeV}$
	motivated by non-observation of the SUSY particles
	and by the "naturalness" arguments.
	Following common practice, we also assume absence of
	a significant cancellation between the two terms in
	definition of $\Lambda_{\alpha}$ given in Eq.
\rf{vec-again}.
	Thus, we come up with the following constraints
\ba{const_num}
\frac{|\mu_{e}|, |\lg\tilde\nu_{e}\rg|}
{\sqrt{(M_{_{\rm SUSY}}/100~\mbox{GeV})}} \leq 90\ \mbox{KeV},\ \ \
76 \ \mbox{KeV} \leq \frac{|\mu_{\mu,\tau}|, |\lg\tilde\nu_{\mu,\tau}\rg|}
{\sqrt{(M_{_{\rm SUSY}}/100~ \mbox{GeV})}}
\leq 276 \ \mbox{KeV}.
\ea
	These constraints apply for all the quantities
	$\mu_{\alpha}$, $\lg\tilde\nu_{\alpha}\rg$  separated
	by commas in the above formulas.
	To our knowledge these stringent limits have not been previously
	derived in the literature. One can find in the published
	papers only those constraints which involve the combinations of
	the 1st and the 2nd generation bilinear \rp parameters
\cite{Now} or contain only the 3rd generation ones
\cite{Mukh}.
	The constraints from Refs.
\cite{Now},
\cite{Mukh} are significantly weaker
	(by several orders) than those in Eq. \rf{const_num}.
	Recent study in Ref. \cite{sugra} concentrate mostly
	on a possibility of simultaneous solution of all the three known
	neutrino anomalies in the GUT constrained \rp MSSM scenarios rather
	than on extraction of constraints on the low energy parameters like
	in Eqs.
\rf{eff-const} and \rf{const_num}.

	It is interesting to note that in the \rp MSSM the generation
	average Majorana neutrino mass is given at
	tree level by the formula:
\ba{nu_ave}
\langle m_{\nu}\rangle = \sum_i m_{\nu_i} \left(V^{(\nu)}_{ei}\right)^2 =
P_{e3} \sqrt{\delta m^2_{\rm atm}} \ .
%m_{\nu} \left(V^{(\nu)}_{e3}\right)^2  = {\cal A}_{e}.
\ea
	Thus, it is expressed in terms of the neutrino
	oscillation parameters.

	As is known the parameter $\langle m_{\nu}\rangle$ plays a clue role
	in the phenomenology of the $\znbb$-decay.
	It determines the classical Majorana neutrino exchange contribution
	to the $\znbb$-decay half-life as
\ba{half-life}
\big[ T_{1/2}^{\znbb}(0^+ \rightarrow 0^+) \big]^{-1}
= G_{01} |{\cal M}_{\nu}|^2 \left|\frac{\langle m_{\nu}\rangle}{m_e}\right|^2.
\ea
	Here $G_{01}$ and ${\cal M}_{\nu}$ are the phase space factor
	and the nuclear matrix element.
	For $\znbb$-decay of ${}^{73}$Ge we have
	$ G_{01} = 7.93\cdot 10^{-15}~{\mbox{years}}^{-1}$
\cite{pa96} and
	${\cal M}_{\nu} = -3.4$
\cite{FKS97}.
	The nuclear matrix element is calculated in the pn-RQRPA approach
\cite{RQRPA}.
	The contribution \rf{half-life} represents a dominant effect
	of the bilinear R-parity violation in $\znbb$-decay
\cite{FKS97}.
	From the best experimental limit on $\znbb$-decay
	half-life of $^{76}$Ge
\cite{hdmo97}
$
T_{1/2}^{{0\nu\beta\beta}}(0^+ \rightarrow 0^+)
\hskip2mm \geq \hskip2mm
1.1 \times 10^{25}~\mbox{years} \ \ (90\% \ \mbox{C.L.})
$
	one obtains $\langle m_{\nu}\rangle \leq 0.50$~eV.
	From the Super-Kamiokande
	and CHOOZ constraints in Eq. \rf{fit}--\rf{Ptau3} we derive
	significantly more stringent limit
\ba{SuperK-znbb}
\langle m_{\nu}\rangle \leq 0.80 \cdot 10^{-2}~\mbox{eV} \ .
\ea
	Thus we conclude that at present the neutrino oscillation
	experiments are more sensitive to the bilinear R-parity
	violation than the $\znbb$-decay experiments.
	This conclusion holds for the whole
	parameter space of the MSSM since both type of the experiments
	place constraints on the same effective parameter ${\cal A}_e$
	which absorbs the dependence on the MSSM parameters.
	To be comparable in sensitivity to this parameter with
	the present neutrino oscillation experiments the
	$\znbb$-experiments have to reach the half-life lower limit
$
T_{1/2}^{{0\nu\beta\beta}} ~\gsim ~10^{29}~\mbox{years}.
$
	This is unrealistic for the running experiments and might only be
	possible for recently proposed large scale germanium detector
\cite{GENIUS}.

%%%%%%%%%%%%%%%%%%%%%%%%%%%%%%%%%%%%%%%%%%%%%%%%%%%%%%%
\vskip10mm
\centerline{\bf Acknowledgements}
\bigskip
S.K. would like to thank the "Deutsche Forschungsgemeinschaft"
for financial support by grant Fa 67/21-1.

%%%%%%%%%%%%%%%%%%%%%%%%%%%

\end{document}